\documentclass[sigconf]{acmart}
\AtBeginDocument{%
  }

\setcopyright{acmlicensed}
\copyrightyear{2018}
\acmYear{2018}
\acmDOI{XXXXXXX.XXXXXXX}
\acmConference[Conference acronym 'XX]{Make sure to enter the correct
  conference title from your rights confirmation email}{June 03--05,
  2018}{Woodstock, NY}
\acmISBN{978-1-4503-XXXX-X/2018/06}

\usepackage{booktabs}
\usepackage{enumitem}
\usepackage{xcolor}
\usepackage{todonotes}
\usepackage{multirow} 

\usepackage{subcaption}
\usepackage{xspace}

\newcounter{box}

\usepackage[framemethod=tikz]{mdframed}

\newcommand{\topicalgraph}{\ensuremath{\mathcal{G}}\xspace}
\newcommand{\topicalnodes}{\ensuremath{\mathcal{V}}\xspace}
\newcommand{\topicaledges}{\ensuremath{\mathcal{E}}\xspace}

\begin{document}



\title{Unexpected Knowledge: Auditing Wikipedia and Grokipedia Search Recommendations}

\author{Erica Coppolillo}
\authornote{Both authors contributed equally to this research.}
\email{erica.coppolillo@unical.it}
\orcid{0000-0002-4670-8157}
\affiliation{%
  \institution{University of Calabria, ICAR-CNR}
  \city{Rende}
  \country{Italy}
}
\affiliation{%
  \institution{University of Southern California}
  \city{Los Angeles}
  \country{California}
}

\author{Simone Mungari}
\authornotemark[1]
\email{simone.mungari@unical.it}
\orcid{0000-0002-0961-4151}
\affiliation{%
  \institution{University of Calabria, ICAR-CNR, Revelis s.r.l.}
  \city{Rende}
  \country{Italy}
}
\affiliation{%
  \institution{University of Southern California}
  \city{Los Angeles}
  \country{California}
}

\begin{abstract}
Encyclopedic knowledge platforms are key gateways through which users explore information online. The recent release of Grokipedia, a fully AI-generated encyclopedia, introduces a new alternative to traditional, well-established platforms like Wikipedia. In this context, search engine mechanisms play an important role in guiding users exploratory paths, yet their behavior across different encyclopedic systems remains underexplored. 
In this work, we address this gap by providing the first comparative analysis of search engine in Wikipedia and Grokipedia.


Using nearly $10,000$ neutral English words and their substrings as queries, we collect over $70,000$ search engine results and examine their semantic alignment, overlap, and topical structure. We find that both platforms frequently generate results that are weakly related to the original query and, in many cases, surface unexpected content starting from innocuous queries. Despite these shared properties, the two systems often produce substantially different recommendation sets for the same query. Through topical annotation and trajectory analysis, we further identify systematic differences in how content categories are surfaced and how search engine results evolve over multiple stages of exploration.

Overall, our findings show that unexpected search engine outcomes are a common feature of {both} the platforms, even though they exhibit discrepancies in terms of topical distribution and query suggestions.  
\end{abstract}

\maketitle

\section{Introduction}

Encyclopedic knowledge platforms play a central role in how users access and explore information on the Web~\cite{Piccardi_Gerlach_West_2024}. A key component of these platforms is the use of search engine mechanisms to guide exploration beyond explicit query results. By suggesting pages, these systems structure users navigation through knowledge spaces. Unlike recommender systems in social media or entertainment domains, encyclopedia-style search engines are typically non-personalized and are commonly assumed to return content that remains close to the user original informational intent. 

Wikipedia, in particular, has become a foundational reference for both everyday information seeking and scholarly research, and has been widely studied as a socio-technical system~\cite{open-access-wikipedia}. At the same time, alternative platforms have emerged that aim to support similar forms of exploratory knowledge access while adopting different technical and organizational design choices. A notable example is represented by \textit{Grokipedia}\footnote{\url{https://en.wikipedia.org/wiki/Grokipedia}}, an AI-based online encyclopedia launched by Elon Musk on October 2025. 
The platform is generated by Grok\footnote{\url{https://grok.com/}
}, a large language model owned by Musk company xAI.

Despite the growing interest in comparing the two platforms in terms of content, references, and structure~\cite{yasseri2025similargrokipediawikipediamultidimensional, article, triedman2025didelonchangecomprehensive}, to the best of our knowledge, no investigation has been performed yet on their respective search-engine systems.

In a effort to rectify this gap, in this paper, we present a systematic comparison of the recommendation behavior of Wikipedia and Grokipedia. Our analysis reveals that, despite their shared goal of supporting exploratory search, both platforms frequently surface recommendations that are unexpected with respect to the starting query. Importantly, such outcomes arise even when users begin with short, neutral, and seemingly innocuous inputs. 

Our \textbf{contributions} can be summarized as follows:

\begin{itemize}[leftmargin=*]
    \item We collect and release the first large-scale dataset aimed to compare the search recommendations provided by Grokipedia and Wikipedia,  consisting of more than $70,000$ recommendations starting from nearly $10,000$ neutral English words.
    \item We examine the semantic alignment between queries and search-engine results, showing that both platforms tend to suggest content that is weakly related to the original query and might lead to unexpected, noxious content.
    \item We extend our analysis to multi-stage recommendation paths, modeling repeated interactions as a topical transition graph. This allows us to assess how unexpected search engine results accumulate over continued exploration, as well as identifying patterns overlooked from single-step recommendations alone.
    
\end{itemize}

The rest of the paper is structured as follows. Section~\ref{sec:related} presents an overview of the current literature in the context of online encyclopedias, search suggestions and navigation modeling. In Section~\ref{sec:setting}, we describe our setting, while in Section~\ref{sec:results} we illustrate the results of our analysis. Section~\ref{sec:conclusions} concludes the paper, addressing its limitations and providing cues for future research directions.

\section{Related Work}
\label{sec:related}

Our work lies at the intersection of search suggestions for encyclopedic platforms, search recommendations, and user navigation modeling. In the following, we first review prior work across these domains and further highlight the gaps that motivate the present study.

\paragraph{\textbf{Grokipedia vs Wikipedia}}
With the recent advent of Grokipedia, growing interest has been raised in comparing the AI-driven encyclopedia with the well-established Wikipedia platform. ~\citet{mehdizadeh2025epistemic} show that Grokipedia citation practices and epistemic foundations diverge substantially from Wikipedia: the former relies less on peer-reviewed academic work and more on user-generated or non-scholarly sources, suggesting a restructuring of epistemic authority in AI-produced content. In \cite{yasseri2025similargrokipediawikipediamultidimensional}, the authors compare the two platforms from both textual and structural perspectives. They show that, although Grokipedia articles often exhibit strong semantic alignment with their Wikipedia counterparts, they tend to be longer yet less lexically diverse, contain fewer references per word, and display greater variability in structural organization. This result therefore suggests a shift toward narrative elaboration rather than rigorous sourcing. Another comprehensive analysis is made in \cite{triedman2025didelonchangecomprehensive}, which reveals that, although much of Grokipedi text is highly derivative of Wikipedia content, its citation patterns include a higher frequency of sources considered unreliable by the Wikipedia community, with particular divergence on politically sensitive topics.

\paragraph{\textbf{Search Suggestions}}
Query Auto‑Completion (QAC) has been extensively studied in information retrieval. Early work analyzed user interactions with QAC systems, revealing positional biases and patterns in suggestion acceptance~\cite{mitra2014user}. Controlled user studies using eye‑tracking show that QAC can alter task completion patterns and reduce the number of result pages visited, with strong position biases shaping user attention and choice~\cite{hofmann2014eye}. Similarly,~\citet{lin2023trapped} reveal algorithmic biases in search-engine auto-completion mechanisms, highlighting how suggestion systems can reflect and amplify societal biases, with disadvantaged groups disproportionately receiving negative or biased completions, underscoring the impact of QAC on equity and perception. Other works focus on supervised and personalized models to improved suggestion relevance by incorporating user history, session context, and demographics~\cite{shokouhi2013learning, kannadasan2019personalized}. Neural approaches further enhance QAC by capturing semantic context and handling unseen prefixes~\cite{wang2020efficient, maurya2023trie}. Additionally, intent- and session-aware models have been proposed to diversify and contextualize suggestions~\cite{kharitonov2013intent}.

\paragraph{\textbf{Navigation Behavior}}
User navigation models have been widely studied in several domains~\cite{hazrati2022recommender, bountouridis2019siren, gao2021imitate, marin2016simulating, carrion2017simulating, coppolillo2024relevance, coppolillo2025algorithmic}. In the context of online encyclopedia exploration, \citet{piccardi_rabbithole} investigate how users navigate Wikipedia beyond surface browsing by analyzing large-scale article access and clickstream data to identify extended navigation patterns, which can lead to rabbit holes. The study quantitatively characterizes these long, multi-step reading sessions and shows that users frequently transition beyond popular pages into the long tail of the encyclopedia, highlighting structural and topical differences in how content is consumed and propagated through hyperlink navigation. In the same area, \citet{rodi_search_strategies} examine how people explore Wikipedia by analyzing clickstream data to uncover general navigation strategies across topics. The authors find that users typically begin their information seeking from broadly scoped, semantically general pages and progressively narrow their focus while maintaining increasing semantic coherence in their path. These patterns differ from those observed in goal-oriented navigation tasks, illustrating distinct cognitive and exploratory behaviors in free browsing versus targeted search.
Other studies focus on random walks and behavioral-cloning models providing tools for simulating user navigation. Prior work has compared real navigation to the random-surfer model~\cite{geigl_random_surfer}.
Further,~\citet{zaheer_navigation} show that agents can learn effective navigation strategies on the Wikipedia hyperlink graph by imitating random walk trajectories. The learned policies enable efficient multi-hop navigation and can complement traditional search methods in tasks such as information retrieval and question answering.
These aforementioned approaches motivate simulation-based evaluation of how different search-completion policies affect arrival probabilities on specific pages.
\\
\\
Despite the growing interest in comparing well-established and AI-driven encyclopedic platforms, to the best of our knowledge no prior work has examined Grokipedia and Wikipedia from the perspective of query completion and search engine suggestions. In this work, we move beyond single-step recommendations by modeling a multi-stage process that simulates the sequential exploration of suggestions a user may follow during query completion.

\section{Setting}
\label{sec:setting}

\paragraph{\textbf{Data}} To conduct our analysis, we select a set of neutral words to be used as search queries. Specifically, we adopt the $10{,}000$ most common English words\footnote{\url{https://github.com/first20hours/google-10000-english}}, identified through an n-gram frequency analysis of the Google Trillion Word Corpus\footnote{\url{https://books.google.com/ngrams/info}}, as described in~\cite{brants2007large}. We use the version of the list in which NSFW terms have been removed, in order to ensure sanitized queries. After this filtering step, the final set of initial queries $Q$ consists of approximately $9{,}900$ words. 

To more effectively analyze the recommendations produced by the Grokipedia and Wikipedia search engines, we issue queries using both complete words and their substrings, thereby mimicking typical human query formulation behavior. Also, in many cases, submitting the complete query alone causes the system to recommend the query itself, thereby limiting the diversity of the suggested results. Recommendations are collected by scraping the corresponding encyclopedia home pages, considering the top-5 suggestions for each query. Overall, this process yields more than $70{,}000$ distinct recommendations, starting from approximately $20{,}000$ distinct queries (including substrings).

\paragraph{\textbf{Reproducibility}} All analyses are conducted using Python, specifically leveraging the package \texttt{selenium}\footnote{\url{https://selenium-python.readthedocs.io/}} to retrieve search engine results from Wikipedia and Grokipedia (version $0.1$). The retrieved dataset and an anonymized version of our code are publicly available to fully ensure reproducibility\footnote{\url{https://anonymous.4open.science/r/Grokipedia_Wikipedia_search_engine-22B5}}.

\section{Analysis}
\label{sec:results}
Our work tackles the following research questions:

\begin{itemize}[leftmargin=0.8cm]
    \item[\textbf{RQ1}:] How do Grokipedia and Wikipedia compare in terms of semantic alignment between queries and search recommendations?
    \item[\textbf{RQ2}:] What is the topical distribution of the recommendations provided by the two platforms?
    \item[\textbf{RQ3}:] Can we detect variations when multi-stage recommendations are considered? 
\end{itemize}

We address each research question in the dedicated paragraphs below.

\begin{figure}[!ht]
\centering
\includegraphics[width=\columnwidth]{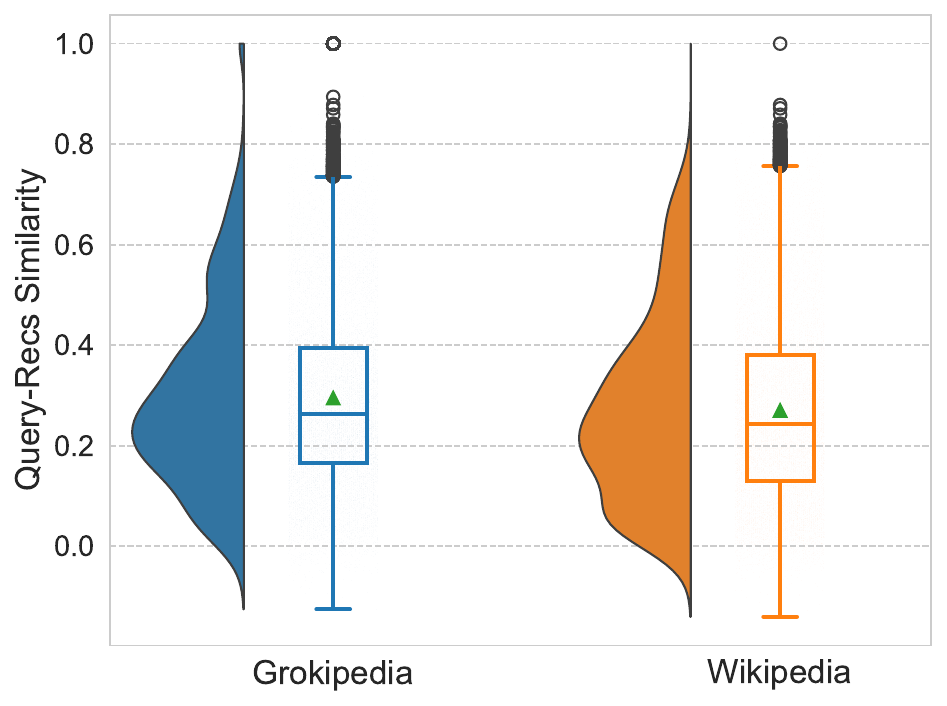}
\caption{Cosine similarity between queries and recommendations provided by Grokipedia and Wikipedia.}
\label{fig:query-recs-similarity}
\end{figure}

\paragraph{\textbf{Query–Recommendation Alignment \rm{\textbf{(RQ1)}}}}
We first assess the semantic alignment between each query and the corresponding recommendations returned by the two search platforms. Query substrings are excluded from this analysis, as they are not relevant to the assessment of semantic alignment. We employ Word2Vec~\cite{mikolov2013efficientestimationwordrepresentations} to generate word embeddings and compute the average cosine similarity between the embedding of the query and those of its recommendations. The results are visualized in Figure~\ref{fig:query-recs-similarity}.

We found that, on average, both platforms provide recommendations with relatively low semantic similarity to the original query. Nevertheless, Grokipedia produces suggestions that are more closely aligned with the searched query ($\mu = 0.30$) than those provided by Wikipedia ($\mu = 0.27$). This difference is statistically significant according to the Mann–Whitney U test~\cite{ROUSSEAUX2013893} ($p < .001$).

\begin{figure}[!ht]
\centering
\includegraphics[width=\columnwidth]{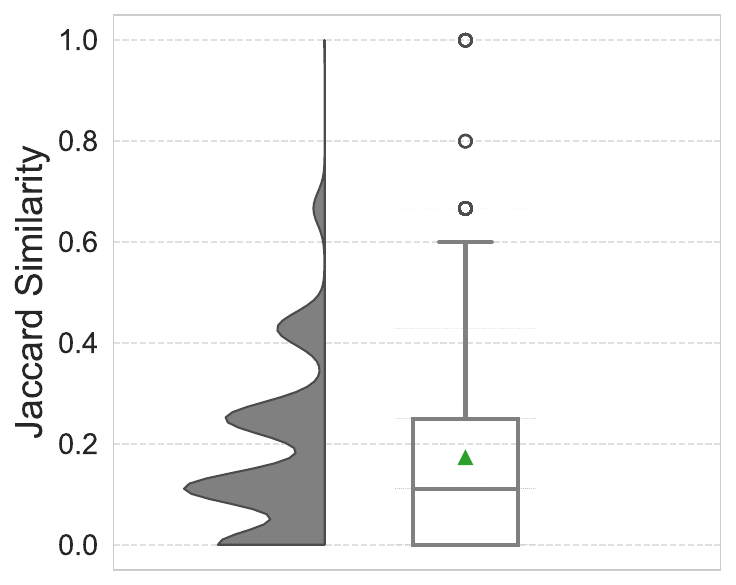}
\caption{Jaccard similarity between Grokipedia and Wikipedia recommendation sets.}
\label{fig:jaccard_similarity}
\end{figure}

Next, we compute the Jaccard similarity~\cite{KANNAN201663} between the recommendation lists generated by the two platforms for each query. The resulting distribution is shown in Figure~\ref{fig:jaccard_similarity}. We observe that the two search engines typically return markedly different sets of recommendations for the same query ($\mu \simeq 0.17, \sigma \simeq 0.16$). This finding motivates a deeper investigation into the nature of the recommended content from a topical perspective.

\paragraph{\textbf{Topical Analysis \rm{\textbf{(RQ2)}}}} We further aim to assess the topic distribution induced by Grokipedia and Wikipedia. To this end, we define a function $f$ which maps a content (either a query or the corresponding search engine suggestion) to one of the following categories: \emph{entertainment, economics, science, law, geography, conspiracy, extremism, adult}, and \emph{other}. 

We instantiate $f$ by means of \texttt{Gemini-2.0-flash}\footnote{\url{https://docs.cloud.google.com/vertex-ai/generative-ai/docs/models/gemini/2-0-flash}}~\cite{google_gemini_update_2024}, a state-of-the-art large language model released by Google in 2024, which demonstrates strong improvements in reasoning and factuality. The model is prompted to assign a semantic topic to each provided content, choosing among the aforesaid categories. The specific prompt is reported in Box~\ref{box:prompt}.

\mdfsetup{skipabove=5pt,skipbelow=5pt}
\refstepcounter{box}\label{box:prompt}
\begin{mdframed}[backgroundcolor=white!10,linecolor=gray!60!,roundcorner=0pt,linewidth=1pt,
rightline=false,
leftline=false] 
\begingroup
\fontsize{8.5pt}{10.5pt}\selectfont
\textbf{Box 1: Gemini System Prompt}
\\
\textit{Determine the semantic topic of the provided word(s).}
\\
\\
\textit{Your answer must fall in the following options:\\
\\
Entertainment, Economics, Science, Law, \\Geography, Conspiracy, Extremism, Adult, Other.}
\endgroup
\end{mdframed}

To assess the robustness of this automatic classification, we randomly sample $100$ words (queries or recommendations) and manually annotate them, reaching complete consensus among the authors. We then compare the model predictions with the human annotations, obtaining an inter-agreement of $88\%$. Notably, most of the misclassifications concern movies or books generically tagged as \textit{other}.

\begin{figure}[!ht]
\centering
\includegraphics[width=\linewidth]{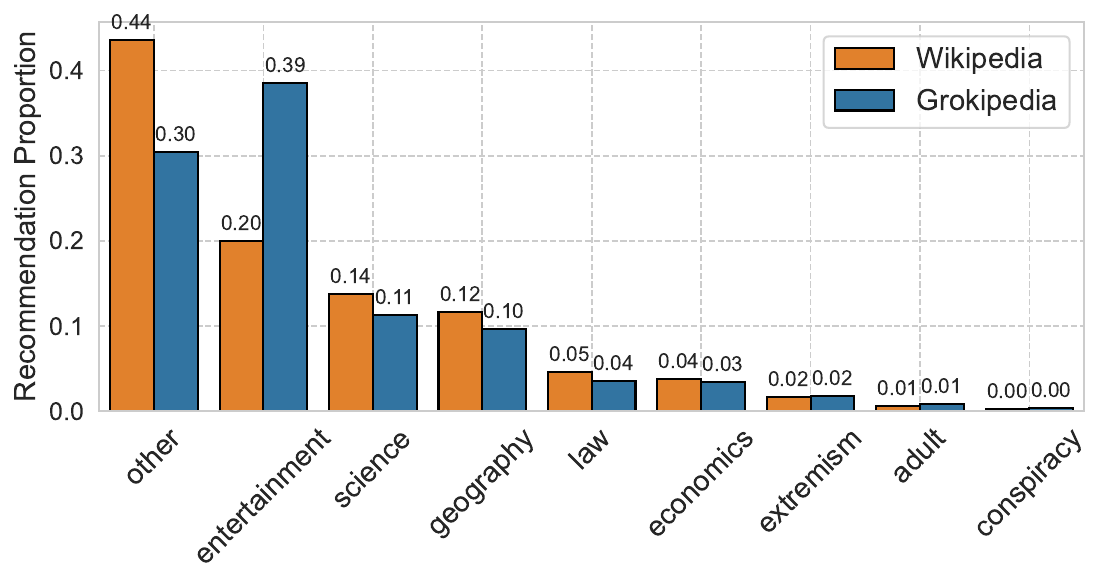}
\caption{Topical distribution of recommendations provided by Grokipedia and Wikipedia.}
\label{fig:recommendation-proportion}
\end{figure}

Figure~\ref{fig:recommendation-proportion} reports the topical distribution of recommendations generated by Grokipedia and Wikipedia.

\begin{figure*}[!ht]
\centering
\begin{subfigure}[b]{0.33\linewidth}
\includegraphics[width=\linewidth]{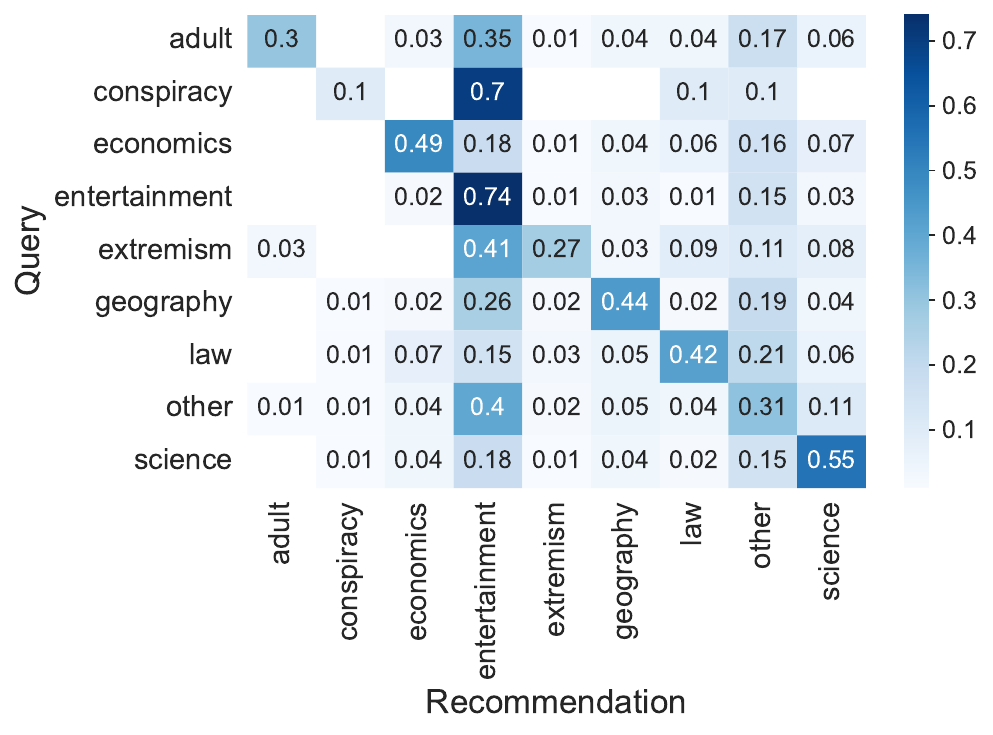}
\caption{Grokipedia\\\textcolor{white}{ciao}}
\label{fig:grokipedia-heatmap}
\end{subfigure}
\begin{subfigure}[b]{0.33\linewidth}
\includegraphics[width=\linewidth]{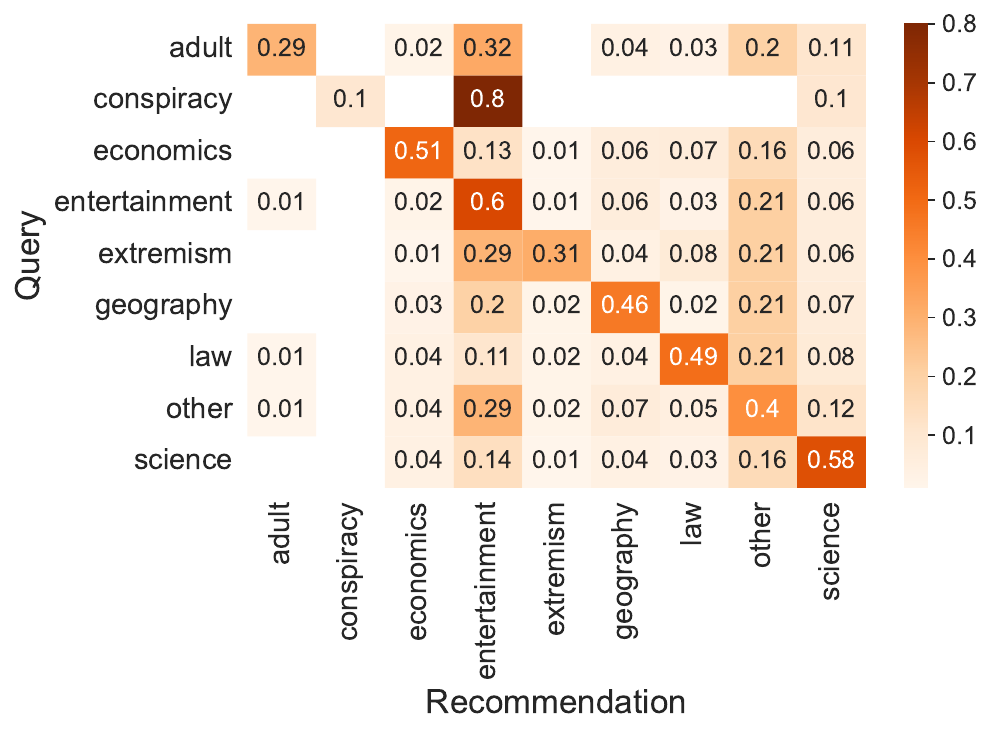}
\caption{Wikipedia\\\textcolor{white}{ciao}}
\label{fig:wikipedia-heatmap}
\end{subfigure}
\begin{subfigure}[b]{0.33\linewidth}
\includegraphics[width=\linewidth]{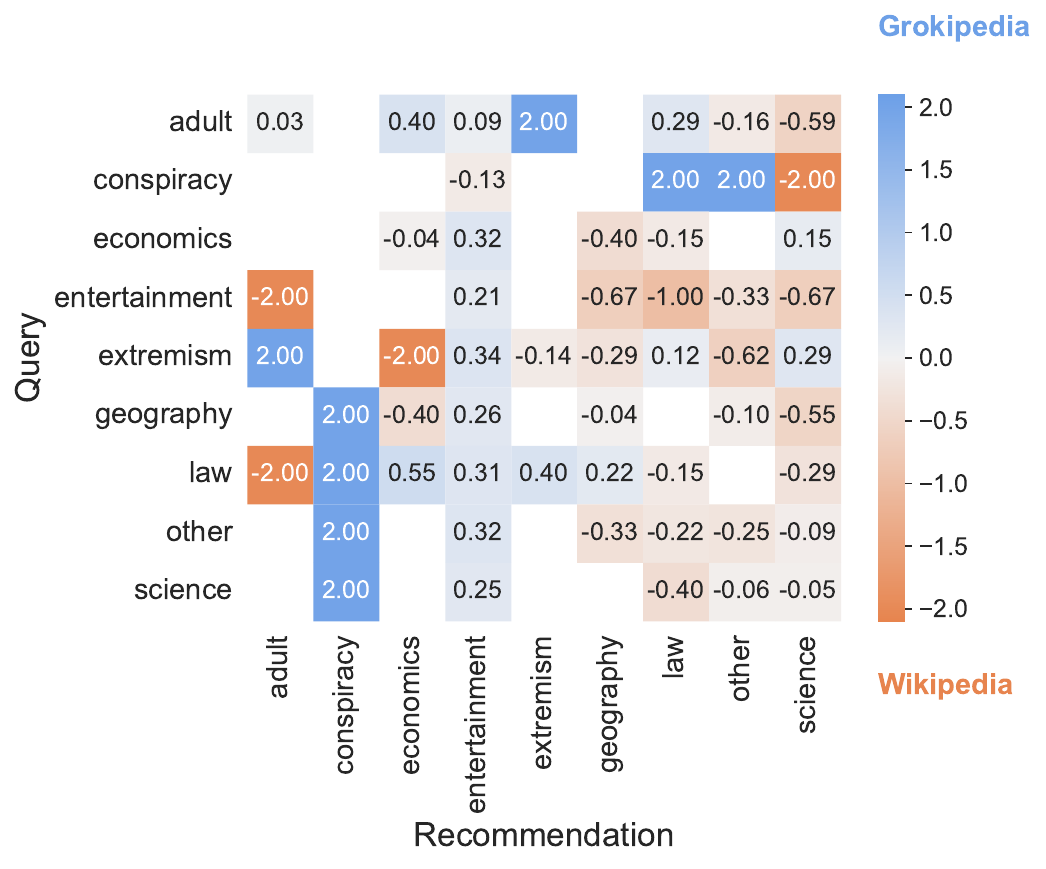}
\caption{Grokipedia - Wikipedia\\(Relative change)}
\label{fig:relative-change-heatmap}
\end{subfigure}
\caption{Proportion of search engine recommendations with a given topic (Recommendation), generated from a given query topic (Query). Blank cells indicate $0$-values.}
\label{fig:topical-heatmap}
\end{figure*}

\begin{table*}[!ht]
\caption{Top-3 recommendations classified as \textit{adult}, \textit{conspiracy}, or \textit{extremism} provided by Grokipedia and Wikipedia. Query columns indicate the shortest distinct subquery(-ies) leading to the recommendation.}
\label{tab:top-3-recommendations}
\resizebox{\linewidth}{!}{
\begin{tabular}{ccllll}
\toprule
{\textbf{Topic}} & {\textbf{Rank}}  & \multicolumn{2}{c}{\textbf{Grokipedia}} & \multicolumn{2}{c}{\textbf{Wikipedia}} \\
\cmidrule(lr){3-4}
\cmidrule(lr){5-6}
 & & \textit{Top Recommendations} & \textit{Query} & \textit{Top Recommendations} & \textit{Query} \\
 \midrule
\multirow{3}{*}{adult}  & 1 & missionary position & pos, mis & attachment in adults & attach \\
 & 2 & sexual practices between women & betw, women, prac & celebrity sex tape & cele \\
 & 3 & human sexual activity & acti, hum, sexu & subsequent pleasures & subseq \\
 \cmidrule(lr){2-6}
\multirow{3}{*}{conspiracy}  & 1 & malaysia airlines flight 370 & malays, fl, ai & recruitment of spies & recru \\
 & 2 & russian sleep experiment & expe, sle & operation gladio & oper \\
 & 3 & june 1962 alcatraz escape attempt & june, atte, esc & continental freemasonry & contine \\
  \cmidrule(lr){2-6}
\multirow{3}{*}{extremism}  & 1 & columbine high school massacre & colu, hig, sch, mas & responsibility for the september 11 attacks & resp \\
 & 2 & heaven's gate (religious group) & rel, gat, heav, grou & involvement of northeast indian insurgents in the myanmar conflict & invol \\
 & 3 & auschwitz concentration camp & camp, conc & responsibility for the holocaust & respo \\
\bottomrule
\end{tabular}
}
\end{table*}

Overall, the two platforms exhibit high degree of alignment in the topic distribution of the search engine results. However, Grokipedia tends to suggest a larger proportion of content classified as {entertainment}, whereas Wikipedia more frequently recommends generic content labeled as {other}. Notably, all categories we identify as \emph{noxious} (i.e., {extremism, adult,} and {conspiracy}) are among the least frequently recommended by both platforms.

We then conduct a more fine-grained analysis of topical \emph{trajectories}, examining how the topic of a query relates to the topics of its recommendations.

Figure~\ref{fig:topical-heatmap} summarizes the results. Figures~\ref{fig:grokipedia-heatmap} and~\ref{fig:wikipedia-heatmap} show the proportion of recommendations of each topic generated from a given query topic on Grokipedia and Wikipedia, respectively. Figure~\ref{fig:relative-change-heatmap} further reports the relative change $\Delta$:
\begin{equation}
\Delta = \frac{p_G - p_W}{(p_G + p_W)/2}
\end{equation}
where $p_G$ and $p_W$ denote the proportions of recommendations produced by Grokipedia and Wikipedia, respectively. In our setting, $\Delta \in [-2, 2]$, where $2$ indicates that a recommendation topic exclusively appears in Grokipedia, and $-2$ indicates the opposite.

Both platforms predominantly recommend content that is topically aligned with the original query, as reflected by the darker diagonal cells. However, the relative comparison reveals notable differences. Wikipedia consistently recommends a higher proportion of adult content when starting from {entertainment} and {law} queries. In contrast, Grokipedia systematically recommends more content classified as {conspiracy} when starting from non-conspiracy queries (e.g., {geography, law, science,} and {other}). Moreover, Grokipedia recommends extremist content less frequently than Wikipedia when starting from extremist queries ($\Delta = -0.14$), but more frequently when starting from {law} queries ($\Delta = 0.40$).

We additionally report in Table~\ref{tab:top-3-recommendations} the top-3 most frequent search engine results classified as {adult}, {conspiracy}, or {extremism} for each platform. We also report the shortest subquery which led to that page. For instance, the reported subquery \textit{malays} denotes that the set of queries \textit{malays}, \textit{malaysi}, and \textit{malaysia} led to the recommendation \textit{malaysiia airlines flight 370}.

Interestingly, both platforms suggest unexpected and seemingly unrelated content starting from innocuous queries. For example, the top adult recommendation provided by Grokipedia is \emph{missionary position}, retrieved from the query \emph{pos} (and its superstrings such as \emph{posi}, \textit{posit}, \dots, \emph{position}). Similarly, Wikipedia suggests \emph{responsibility for the September 11 attacks} when prompted with the substring \emph{resp}. These observations suggest that both platforms may guide users toward unexpected content starting from semantically distant queries. This motivates a further investigation of recommendation behavior across multiple recommendation stages.

\paragraph{\textbf{Multi-stage Recommendations \rm{\textbf{(RQ3)}}}}

\begin{figure*}[!ht]
    \centering
    \begin{subfigure}[b]{0.49\linewidth}
        \includegraphics[width=\linewidth]{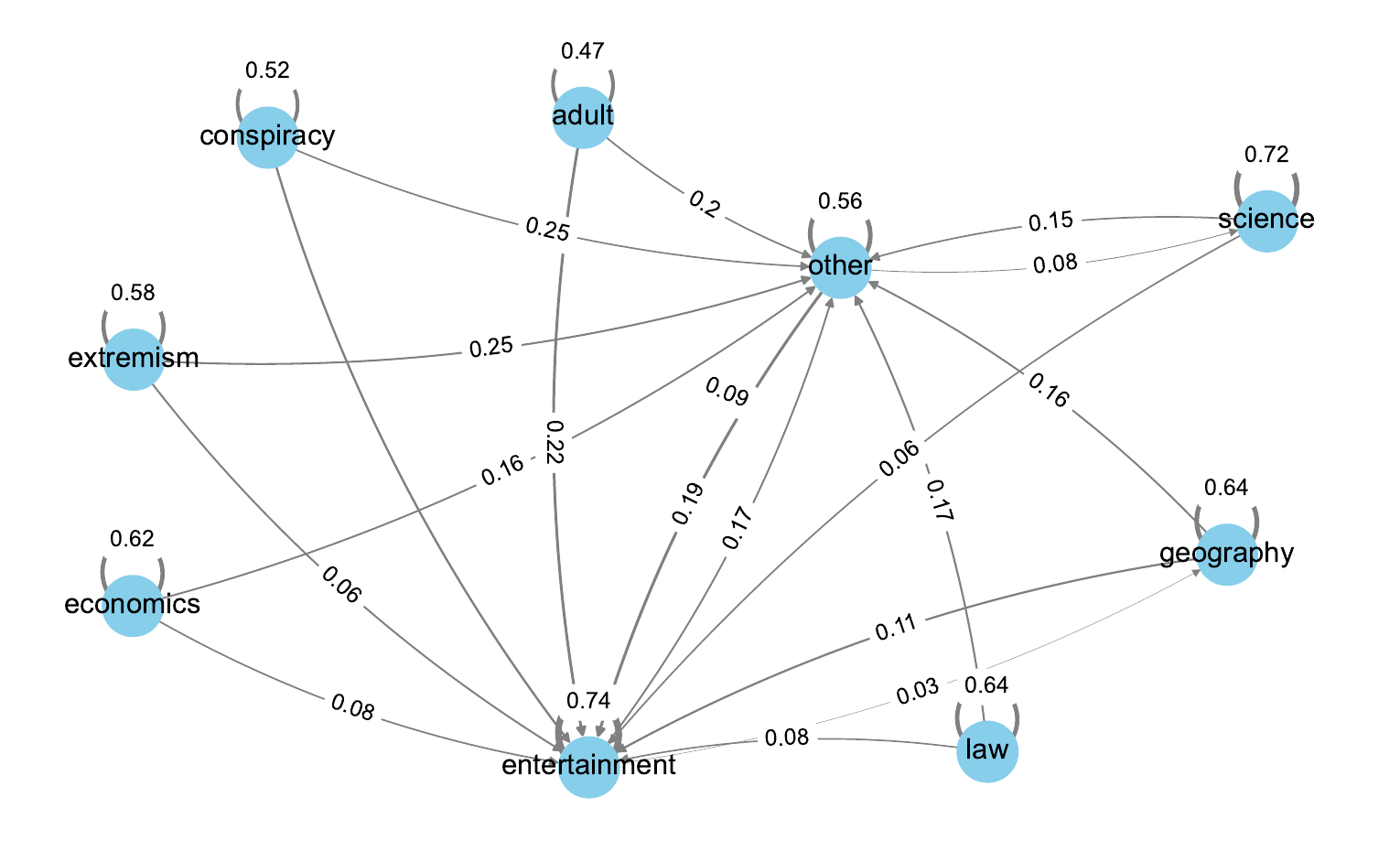}
        \caption{Grokipedia}
    \end{subfigure}
    \begin{subfigure}[b]{0.49\linewidth}
        \includegraphics[width=\linewidth]{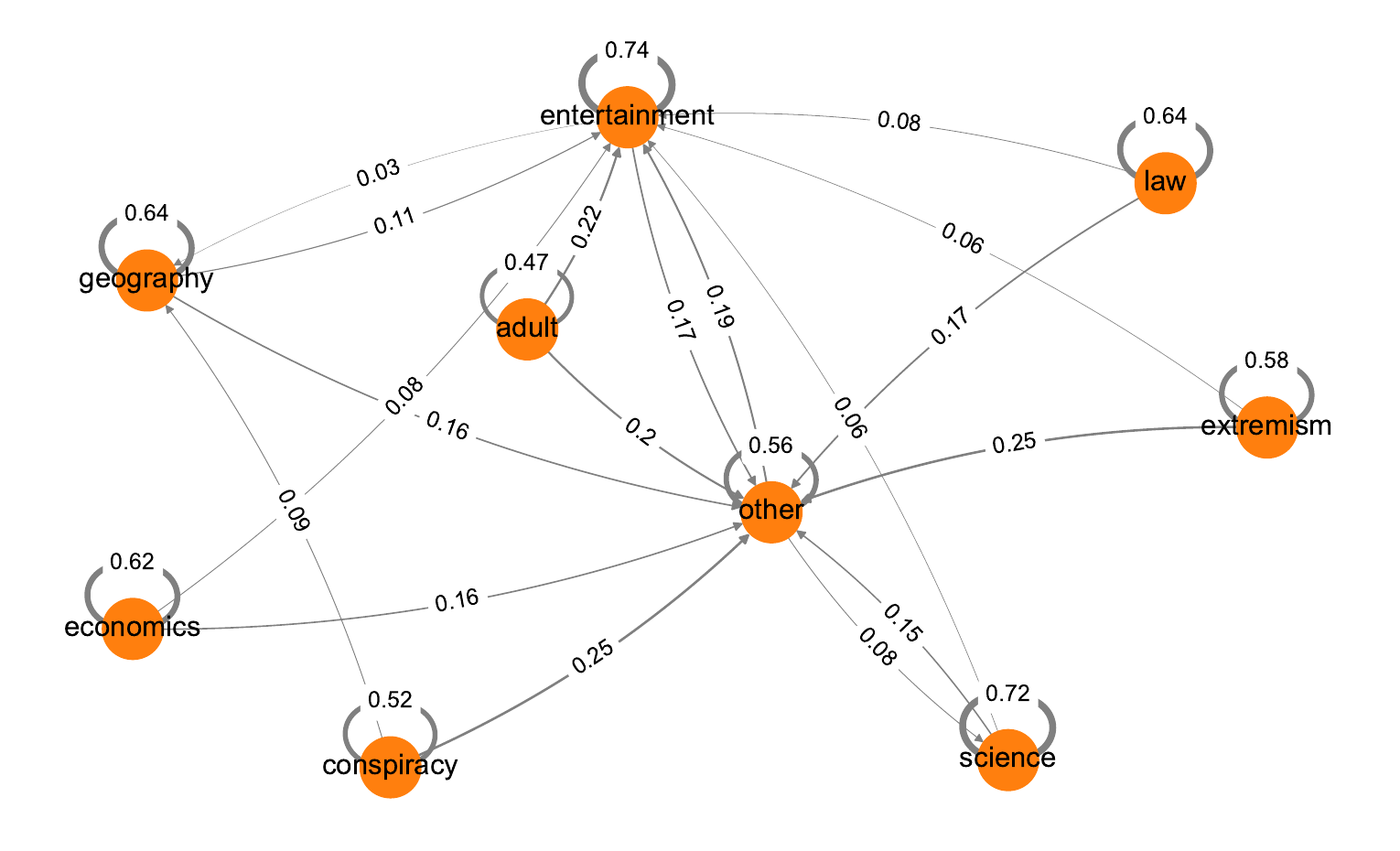}
        \caption{Wikipedia}
    \end{subfigure}
    \caption{Topical multi-stage recommendation graphs. Nodes represent the encountered topics, while edges indicate transitions between topics. Edge weights show the probability of transition from a topic to another. To ensure readability, only the 3-most common topical transitions are visualized.}
    \label{fig:topical-graph}
\end{figure*}

We now examine how topical distributions evolve across multiple stages of recommendations. To this end, we construct a directed probabilistic \emph{topical graph} $\topicalgraph = (\topicalnodes, \topicaledges, \mathcal{P})$, where $\topicalnodes$ is the set of encountered topics.
Each edge $(v_1, v_2) \in \topicaledges$ represents a possible transition between nodes (i.e., topics) and is associated with the probability $p(v_2|v_1)\in [0,1]$.

To construct the topical graph \topicalgraph, we first employ a depth-first search (DFS) strategy to systematically explore multi-stage recommendation paths. Specifically, for each query $q \in Q$ and for each substring obtained by incrementally revealing the characters of $q$, we recursively apply the DFS algorithm to retrieve search engine recommendations up to a maximum depth of $3$. At each recursion level, the top-1 recommendation $r$ (possibly different from the corresponding query) is used as input for the subsequent level, thereby simulating the sequential exploration of suggestions that a user may follow during the query completion process\footnote{For each step, we limit to a \textit{single} recommendation given the intrinsically exponential nature of the process.}. In practice, at each step, a new edge $(f(q), f(r))$ is added to \topicaledges, where $f(q)$ is the topic of the (original) query $q$ and $f(r)$ is the topic of the top-1 new retrieved recommendation (either produced by $q$ itself or by one of its substrings). This procedure enables us to capture topic transitions across multiple recommendation stages, which are then aggregated to estimate the transition probabilities encoded in \topicalgraph. 
Notably, we adopt this procedure as a proxy for the ``Related pages'' feature available in Wikipedia, as no equivalent functionality is currently available in Grokipedia.
A visualization of the topical graphs of Grokipedia and Wikipedia is depicted in Figure~\ref{fig:topical-graph}. 

\begin{figure*}[!ht]
\centering
\begin{subfigure}[b]{0.33\linewidth}
\includegraphics[width=\linewidth]{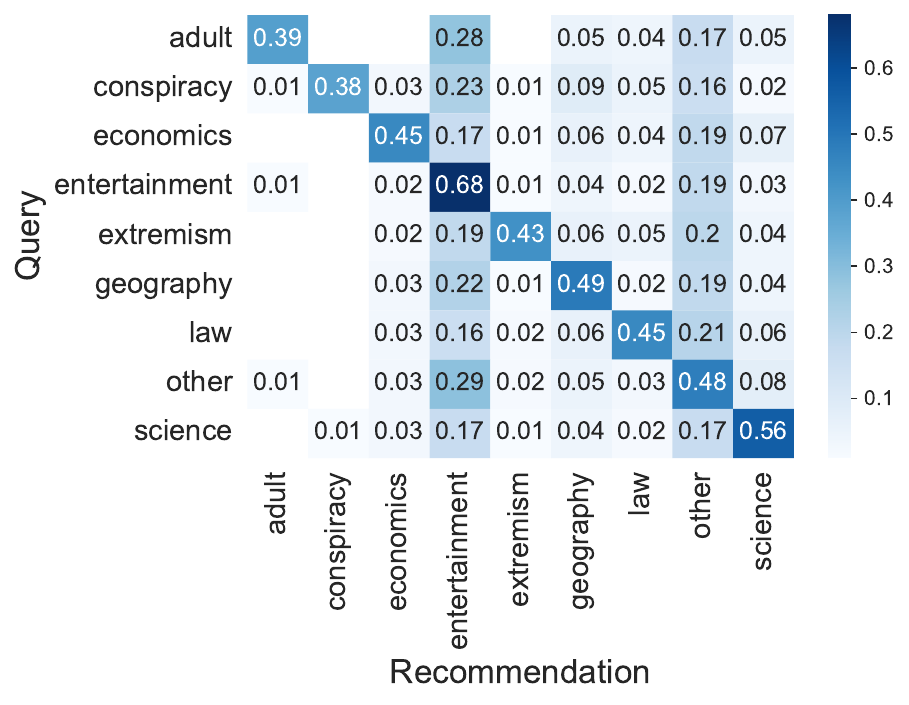}
\caption{Grokipedia\\\textcolor{white}{ciao}}
\label{fig:graph-grokipedia-heatmap}
\end{subfigure}
\begin{subfigure}[b]{0.33\linewidth}
\includegraphics[width=\linewidth]{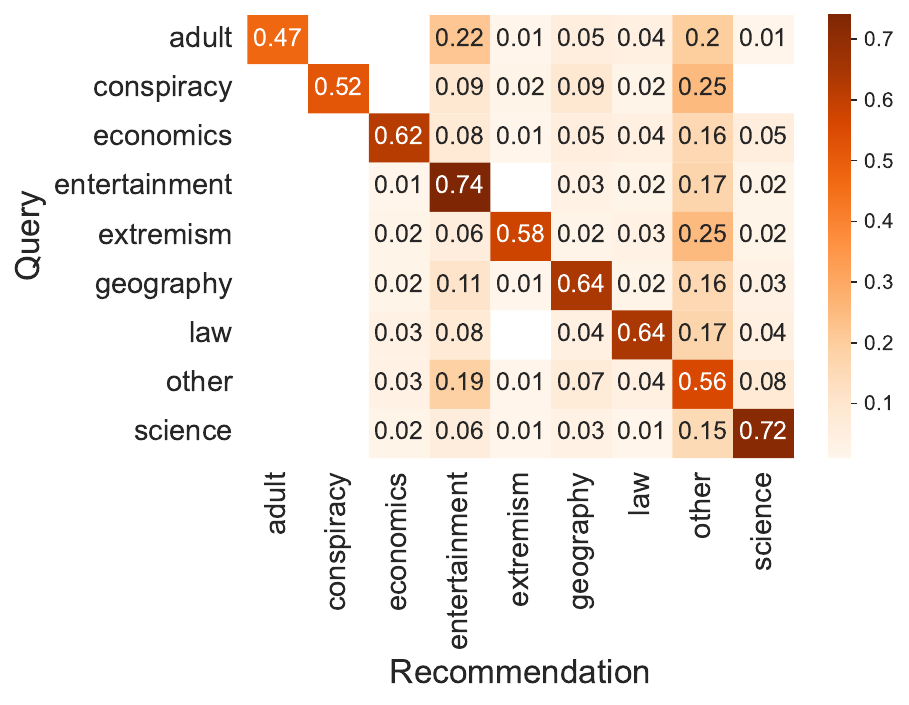}
\caption{Wikipedia\\\textcolor{white}{ciao}}
\label{fig:graph-wikipedia-heatmap}
\end{subfigure}
\begin{subfigure}[b]{0.33\linewidth}
\includegraphics[width=\linewidth]{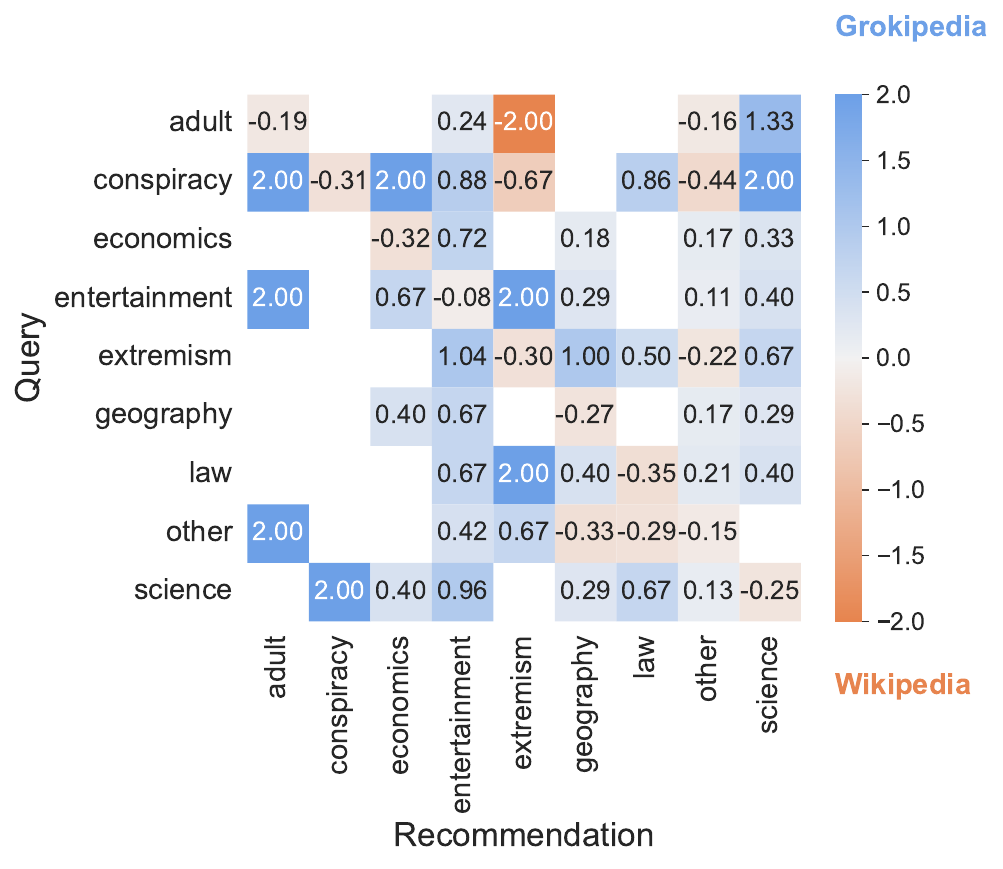}
\caption{Grokipedia - Wikipedia\\(Relative change)}
\label{fig:graph-relative-change-heatmap}
\end{subfigure}
\caption{Proportion of search engine recommendations with a given topic (Recommendation), generated from a given query topic (Query), computed over the multi-stage recommendation graph $\topicalgraph$. Blank cells indicate $0$-values.}
\label{fig:graph-topical-heatmap}
\end{figure*}

The results concerning the evolution of topical distribution (from query to recommendations) are further shown in Figure~\ref{fig:graph-topical-heatmap}. As in the single-stage analysis, Figures~\ref{fig:graph-grokipedia-heatmap} and~\ref{fig:graph-wikipedia-heatmap} report the proportions of recommendations by topic for Grokipedia and Wikipedia, respectively, while Figure~\ref{fig:graph-relative-change-heatmap} shows the relative change $\Delta$.
Compared to the single-stage setting, the multi-stage analysis amplifies topical polarization on both platforms, as indicated by the darker diagonal elements. At the same time, new relative patterns emerge. Wikipedia more consistently recommends extremist content starting from {adult} ($\Delta = -2$) and {conspiracy} ($\Delta = -0.67$) queries. Conversely, Grokipedia is the only platform that recommends adult content starting from {conspiracy}, {entertainment}, and {adult} queries; extremist content starting from {geography} and {law}; and scientific content starting from {conspiracy} queries.

Finally, Table~\ref{tab:top-3-recommendations-graph} reports the top-3 recommendations classified as {adult}, {conspiracy}, or {extremism} when considering the full recommendation graph $\topicalgraph$. Again, several unexpected results emerge. For instance, Grokipedia suggests \emph{coital alignment technique} when queried with \emph{techni}, \textit{coil}, and \textit{alig}, while Wikipedia recommends \emph{donald trump access hollywood tape} starting from the substring \emph{donal}. Notably, both platforms return identical top recommendations in several cases, including \emph{sexual intercourse} (top-1 adult recommendation from the query \emph{sex}), \emph{january 6 united states capitol attack} (top-1 extremism recommendation from \emph{jan}), and \emph{provisional irish republican army} (top-3 for Grokipedia and top-2 for Wikipedia from \emph{prov}).

Although Grokipedia produces unexpected recommendations when starting from a higher number of subqueries, our multi-stage analysis confirms that \textit{both} encyclopedic platforms suggest noxious pages in response to queries that appear unrelated and non-suspect.

\begin{table*}[!ht]
\caption{Top-3 noxious recommendations computed on the multi-stage recommendation graph, provided by Grokipedia and Wikipedia. Query columns indicate the shortest distinct subquery(-ies) leading to the recommendation.}
\label{tab:top-3-recommendations-graph}
\resizebox{\linewidth}{!}{
\begin{tabular}{ccllll}
\toprule
{\textbf{Topic}} & {\textbf{Rank}}  & \multicolumn{2}{c}{\textbf{Grokipedia}} & \multicolumn{2}{c}{\textbf{Wikipedia}} \\
\cmidrule(lr){3-4}
\cmidrule(lr){5-6}
 & & \textit{Top Recommendations} & \textit{Query} & \textit{Top Recommendations} & \textit{Query} \\
 \midrule
\multirow{3}{*}{adult}  & 1 & sexual intercourse & sex & sexual intercourse & sex \\
 & 2 & coital alignment technique & techni, coi, alig & donald trump access hollywood tape & donal \\
 & 3 & inappropriateness & inap & lingerie & linge \\
 \cmidrule(lr){2-6}
\multirow{3}{*}{conspiracy}  & 1 & death of diana, princess of wales & deat, wales, diana & september 11 attacks advance-knowledge conspiracy theories & septe \\
 & 2 & malaysia airlines flight 370 & ai, fl, malays & the protocols of the elders of zion & prot, the p\\
 & 3 & hunter biden laptop controversy & contr, lap, hunter & knights templar & kn \\
 \cmidrule(lr){2-6}
\multirow{3}{*}{extremism}  & 1 & january 6 united states capitol attack & jan, cap, att, united & january 6 united states capitol attack & jan \\
 & 2 & columbine high school massacre & hig, sch, col, mas & provisional irish republican army & prov \\
 & 3 & provisional irish republican army & iri, republica, prov & provisional irish republican army campaign & prov  \\

\bottomrule
\end{tabular}
}
\end{table*}

\section{Conclusions}
\label{sec:conclusions}
In this paper, we presented the first large-scale comparative analysis of the \textit{search recommendation} behavior of Wikipedia and Grokipedia, two encyclopedia platforms designed to support exploratory access to online knowledge. By auditing their recommendations using a broad set of neutral queries and query substrings, we examined how these platforms differ and overlap in terms of semantic alignment and topical composition, in both single- and multi-stage exploration dynamics.

Our results show that both platforms frequently surface recommendations that are only weakly semantically related to the original query and, in many cases, lead to unexpected content starting from innocuous inputs. At the same time, Wikipedia and Grokipedia often produce substantially different recommendation sets for the same query, revealing that platform-specific design and implementation choices meaningfully shape users exploratory pathways.

Through topical analysis, we further showed that recommendations tend to remain broadly aligned with the query topic, yet transitions across topics, including toward \textit{sensitive} categories (i.e., adult, conspiracy, extremism), do occur. Extending the analysis to multi-stage recommendation paths highlighted that such patterns persist and can become more pronounced over repeated interactions, reinforcing the importance of considering recommendation dynamics beyond single-step outputs.

Taken together, these findings challenge the common assumption that encyclopedic search engines reliably constrain exploration to semantically proximate or expected content. More broadly, our results show that unexpected dynamics can emerge across both human-curated platforms, such as Wikipedia, and AI-driven encyclopedic systems, such as Grokipedia.
\paragraph{\textbf{Limitations}} This study has some limitations that should be considered when interpreting the results. First, our analysis focuses on a fixed snapshot of recommendation behavior and does not account for temporal variation. Both Wikipedia and Grokipedia may update their recommendation mechanisms over time, and future changes could alter the observed patterns. Second, while we use a large and diverse set of neutral English words, our query set does not capture all possible forms of user interaction. In particular, we do not consider longer natural-language queries, multilingual inputs, or queries informed by prior context, all of which may influence the search engine result. Third, our topical classification relies on a large language model, which, despite demonstrating strong agreement with human annotations, may introduce classification errors or biases. Although we validated the model on a manually annotated sample, topical boundaries remain inherently subjective, and alternative taxonomies could yield different insights.
Finally, our study examines recommendation outputs but does not directly model user behavior or user perception. While we identify unexpected recommendations from innocuous queries, we do not assess how users interpret, engage with, or are influenced by such content in real-world settings.

\paragraph{\textbf{Future Work}} This work opens several promising directions for future research. First, longitudinal studies could track how recommendation behavior evolves over time, enabling the detection of shifts in platform design, ranking strategies, or content coverage. Such analyses would be particularly valuable for understanding the stability of unexpected recommendation patterns.
Second, future work could extend the comparison to additional knowledge platforms or alternative interfaces to Wikipedia, allowing for a broader characterization of design choices in encyclopedic search systems. Comparative analyses across languages and cultural contexts would further enrich this perspective.
Third, incorporating user-centered methods, such as controlled user studies or interaction logs, would help bridge the gap between observed recommendation patterns and their practical implications for information seeking and understanding. This would allow researchers to assess whether unexpected recommendations are perceived as useful serendipity, confusion, or noise.
Finally, methodological extensions could explore finer-grained semantic or structural representations of recommendation pathways, including graph-based metrics or causal analyses. Such approaches could help disentangle how specific design elements contribute to observed differences between platforms.

\bibliographystyle{ACM-Reference-Format}
\bibliography{bibliography}

\end{document}